\documentclass[10pt]{article}

\usepackage[all]{xy}
\usepackage{amssymb}
\usepackage{amsthm}
\usepackage{amsmath}
\usepackage{newlfont}
\usepackage[T1]{fontenc}
\usepackage{mathrsfs}
\usepackage{vmargin}
\usepackage{endnotes}
\usepackage{graphicx}
\usepackage{verbatim}
\usepackage{float}
\usepackage{multirow}
\usepackage{setspace}
 \usepackage[bf]{caption}
 \usepackage{enumerate}

\usepackage{vmargin}
    \setpapersize{A4}
    \setmarginsrb{20mm}{13mm}{20mm}{15mm}
                 {6mm}{6mm}{12mm}{12mm} 

\usepackage[latin1]{inputenc}
\usepackage[english]{babel}
\usepackage{graphicx}
\numberwithin{equation}{section}

\newtheorem{proposition}{Proposition}

\theoremstyle{definition}

\newcommand{\target}{\overline \sigma}

\newcommand{\R}{\mathbb R}
\newcommand{\E}{\mathbb E}

\newcommand{\I}{1{\hskip -3.4 pt}\hbox{I}}

\newcommand{\Vol}{I_T}

\def\registered{{\ooalign{\hfil\raise .00ex\hbox{\scriptsize R}\hfil\crcr\mathhexbox20D}}}

\begin{document}

\title{\bf{Pricing joint claims on an asset and its realized variance under stochastic volatility models}\footnote{Part of this material was presented at Santander monthly seminar in quantitative finance, London, January 2011.  The author would like to thank William Shaw and Giuseppe Di Graziano for their useful comments.  }}
\author{Lorenzo Torricelli \footnote{ Email: lorenzo.torricelli.11@ucl.ac.uk }\\
Department of Mathematics\\
University College London
}
\maketitle
\hspace{. cm}
\begin{abstract} In a stochastic volatility framework, we find a general pricing equation for the class of payoffs depending on the terminal value of a market asset and its final quadratic variation. This allows a pricing tool for European-style claims paying off at maturity a joint function of the underlying and its realised volatility/variance. We study the solution under different stochastic volatility models, give a formula for the computation of the Delta and Gamma of these claims, and introduce some new interesting payoffs that can be priced through this equation. Numerical results are given and compared to those from plain vanilla derivatives.
\end{abstract}

\smallskip


\smallskip


\section{Introduction}

The interest of markets in volatility derivatives has significantly grown since the late nineties. The reason of such an endorsement by traders and investors is that volatility-linked products allow trading in equity markets without necessarily having to take a position in the underlying, thus avoiding the typical problems associated to delta-hedging.  As a result, a large body of theory on the pricing and hedging of these products has emerged. Neuberger \cite{Neuberger} and Dupire \cite{Dupire} pioneered the classic replication of a variance swap via the log-contract. Since then the model-independent approach to pricing volatility derivatives has been widely developed, most notably during the last decade (e.g. Carr and Lee \cite{CarrLee}). At the same time methods relying on specific models have been proposed; see for instance Matytsin \cite{Matytsin}, Elliott \cite{Elliott} \emph{et al},  Howison \cite{Howison} \emph{et al},  Javaheri \cite{Al-Javaheri} \emph{et al}.

In the years immediately before the last market crisis, new kinds of volatility-related investment paradigms begun to arise. Unlike the \emph{pure volatility} derivatives already present in the market, these new products are based on an equity-linked underlying, and use the future realised volatility (i.e. the averaged standard deviation of the log-returns) of this same asset as an adjusting factor. In a broad sense, the purpose of this construction is to allow a position in the underlying in a classic sense (options, futures) while at the same reducing the inherent volatility risk. The most popular investments of this kind are portfolio management strategies generically known as target volatility indices, consisting in a periodical review of  the underlying exposure according to the realised volatility. The purpose is to reduce the vega effect due to the volatility fluctuations for the duration of the portfolio. Contracts written on these indices are then in principle priced  ``as if'' the volatility it exhibits was constant. This means not only a safer investment but also a much simplified, Black-Scholes alike, pricing framework. Examples of target volatility indices sponsored by investment companies are Russell's Controlled Volatility Indices , JPMorgan's Commodity Target Volatility Indices, Morningstar Ultimate Stock-Pickers
Target Volatility Indexes, DWS NASDAQ-100 Volatility Target Index and BarCap's Revolver.  Also, detailed financial studies of target volatility indices performances are beginning to appear (Chew, \cite{Chew}).

Given the sophisticated nature of the target volatility indices, a first step towards an analytical valuation theory of these instruments entails of course understanding how a ``vanilla'' joint asset and volatility payoff is to be priced.  Such a vanilla  product is the so-called \emph{target volatility option} (TVO). A TVO is an European-type derivative contract that pays off at maturity a \emph{random} fraction of a vanilla European Call, times a \emph{target volatility} parameter $\bar \sigma$ representing the investor's expectation of the future realised volatility. The random fraction considered is the inverse of the realised volatility of the underlying. As one might expect, it turns out that at-the-money, this derivative is approximately priced by a Black Scholes Call option having implied volatility $\bar \sigma$. Introducing a volatility adjustment in the payoff effectively determines the reduction of the price paid for. Whether one will benefit from this while cashing in the payoff ultimately depends on the accuracy of the volatility prediction $\bar \sigma$. 

A study of the TVO and related pricing methodologies has begun in Di Graziano and Torricelli, \cite{Torricelli}. We gave there three distinct pricing methodologies for the TVO, under a stochastic volatility model where no correlation is assumed between the asset and the stochastic volatility driving Brownian components.

 In this paper we develop further some of the ideas of \cite{Torricelli}, and we do so in two different directions. In first place, we remove the assumption of independence between the asset and the stochastic volatility; secondly, we find a general pricing technique valid not only for the TVO payoff, but for \emph{any} sufficiently regular joint asset and volatility payoff. 

The latter contribution, far from being a mere theoretical exercise,  may very well yield to sensible real-life derivative products.   As we shall see, by using a volatility correction, it is possible to modify a European payoff $f$ into a claim $\tilde{f}$, in such a way that taking a position in $\tilde{f}$ will be less costly, and still produce the same payoff as $f$ if some predicted volatility event takes place. If an investor wishes to trade in $f$ and has a strong belief about future volatility, she may choose trading in $\tilde{f}$ instead, eventually being better off if her prediction was correct. 

This naturally brings in the question of finding a way to price the class of general claims $F(x,y)$ of two variables: one representing the asset's terminal value, the other one its realised volatility.  In a general stochastic volatility framework, we find a partial differential equation giving the time-$t$ price of a contract written jointly on an asset as well as on its realised volatility, and solve it by Fourier transform methods.

 More precisely, denoting the quadratic variation of the log-returns of a market asset $S_t$ by \begin{equation}I_t=\int_0^t v_u d_u,\end{equation}  where $v_t$ is the instantaneous variance, we are hereby interested in pricing European-style contingent claims maturing at time $T$ and having the form:

 \begin{equation}\label{payoff}F_T=F(S_T, I_T),\end{equation}

\smallskip

\noindent for some function of two variables $F$. Our fundamental variables will thus be the asset and its quadratic variation. In a continuous stochastic volatility model the statistical realised variance and volatility of $S_t$ are approximated respectively by $I_T/T$ and $\sqrt{I_T/T}$, so that the class (\ref{payoff}) is completely equivalent to that of the \emph{joint asset and realised volatility (variance) claims}.




Although our main result is dependent upon the \emph{choice} and the \emph{calibration} of the dynamics for the stochastic variance, it is \emph{universal} in the sense that works with any sufficiently well-behaved such model.
The possible alternative approach, a \emph{parameter-free} replication pricing like the one advocated by Carr and Lee in \cite{CarrLee}, although certainly not prone to estimation errors, would necessarily rely on the famous formula by Breeden and Litzenberger of \cite{BL}, representing a claim on an asset through a portfolio of European Calls and Puts. In practice, the issue with this technique is that the market may not offer a sufficient range or density of traded strikes, leading to a truncation or discretisation errors in the formula, especially for long maturities.  In such instances a \emph{model-dependent} choice, like the one of this paper, may be preferable.

 Similar versions of our main equation are already present in literature. To our knowledge, Lipton \cite{Lipton} first gave the Fokker-Planck equation corresponding to the log-price version of (\ref{PDE}). Fatone \emph{et al.} \cite{Fatone} gave a solution for such backward equation in the Heston model by means of a Fourier inversion, and then used the arising family of probability densities to obtain the price of \emph{pure volatility} derivatives. Therefore, in their work a double integration is needed for a claim depending on a single state variable. Sepp \cite{Sepp} instead presents (the log-transformation of) equation (\ref{PDE}) for jump diffusions and solves it with a method similar to ours, but then he excludes the price variable from the analysis and reverts to solutions for \emph{pure volatility} derivatives in the Heston model.

 In contrast, we will obtain pricing formulae for claims depending also on the final asset price at expiry for \emph{any} well-behaved stochastic volatility model, while at the same time keeping the integration involved to the minimum. 

 In the spirit of the systematic study by Heath and Schweizer \cite{Heath}, great care has been take in emphasizing a series of sufficient conditions that make the pricing problem mathematically unambiguous.  
 We do this by referring to the classic theory of parabolic equations and SDEs (Friedman \cite{Friedman}, \cite{Friedman2}, Feller \cite{Feller1}, Kunita \cite{Kunita}) which we hope to revive in the context of financial PDEs.

 The solution approach proposed is the natural 2-dimensional extension of the method for pricing derivatives in a stochastic volatility framework introduced by Lewis in \cite{Lewis}. Our \emph{fundamental transform} will be taken with respect to the quadratic variation $I$, besides the log-price $x$. Strikingly, only minor modifications in the final formulae of Lewis are needed, which accounts how powerful his method is. Aside, it is noteworthy that the several-dimension Fourier transform idea can be in principle applied to pricing of claims depending on other kinds of non-traded market factors (e.g. Asian options).

In section 2 we will define our model and derive the pricing equation, which is solved in section 3 together with a derivation for the Greeks. Section 4 shows fundamental transforms for various models and discusses their existence. In section 5 some claims of the form $F(S_T, I_T)$ are introduced, which are then tested numerically in section 6. Technical details are proved in the appendix.

\section{Setting up the equation}

The single asset scenario we are assuming consists of a three factor It\^{o} process $X_t=(S_t, v_t, I_t)$ describing the evolution in time of a \emph{risky asset} $S_t$, its \emph{stochastic instantaneous variance }$v_t$ and its \emph{realised variance} $I_t$. A constant market risk-free rate $r$ exists, and the asset $S_t$ continuously pays to its owner a proportional constant dividend yield $d$. Valuations relying on such a stochastic variance model are clearly unique modulo different choices of a market price of risk, which we hereafter assume to be fixed. This induces a risk-neutral pricing probability measure $\mathbb P$, which is the only one relevant in all that will follow. Under such a law the price $S_t$ will therefore exhibit a log-return rate of $r-d$.   

Let $(\Omega, \mathbb P, \mathcal F,   \{\mathcal F_t\}_{ t  \geq 0})$ be a market filtration satisfying the usual conditions, and let $W^1_t$, $W^2_t$ be two $\mathcal F_t$-adapted Brownian motions having correlation $\rho  t$. The underlying diffusion $X_t$ is assumed to be of the form:

\begin{equation}\label{D}\tag{D}
\left \{
\begin{array}{l}
dS_t= (r-d) S_t dt + \sqrt{v_t} S_t d W^1_t\\
d v_t= \alpha (t,v_t) dt+ \beta(t,v_t) dW^2_t \\
d I_t=v_t dt.
\\\end{array}
\right.
\end{equation}

For our purposes we are interested in the behaviour of this process in a finite time range $[0,T]$. We assume the coefficients $\alpha(t,x) : \mathbb R^2_+ \rightarrow \mathbb R$, $\beta(t,x) : \mathbb R^2_+ \rightarrow \mathbb R_+$ to be locally Lipschitz-continuous in $x$, uniformly in $t$;  that is, for all compact sets $K \subset \mathbb R_+$, $\exists C_K>0$ such that:
\begin{equation}\label{LL}\tag{LL}
\begin{array}{cc}
\displaystyle \sup_{0 \leq t \leq T} |\alpha(t,x)-\alpha(t,y)|+\sup_{0 \leq t \leq T}|\beta(t,x)-\beta(t,y)| < C_K |x-y|, &  \forall x,y \in K.\\
\end{array}
\end{equation}

  In general, (\ref{LL}) is  sufficient to ensure that a unique strong solution to (\ref{D}) exists only up to a random exit time of $\mathbb R^3_+$. To obtain an everywhere well defined solution we must impose the (in the words of Feller) \emph{natural boundary conditions}:
\begin{equation}\label{NB}\tag{NB}\begin{array}{cc}\displaystyle{\mathbb P_{x,t}\left(\sup_{t \leq s \leq T} v_s=+ \infty \right)=\mathbb P_{x,t}\left(\inf_{t \leq s \leq T} v_s=0 \right)=0}, & \forall x \in \mathbb R_+, t \in [0,T).\end{array} \end{equation}

We also say that the $0$ and $+ \infty$ boundaries for the variance process must not be \emph{attainable}. Most of the commonly used models for stochastic volatility satisfy (\ref{LL}) but not necessarily (\ref{NB}) for every possible choice of parameters.  In practice, failure to meet (\ref{NB}) has to be interpreted as the possibility of volatility explosions or volatility vanishing. When this happens multiple solutions for $X_t$ are possible. However, while treating specific models we will impose suitable conditions under which (\ref{NB}) is satisfied, so that $X_t$ will always be unique.

 Now suppose we want to trade a derivative that pays off at the maturity date $T$ a certain function of two variables: the underlying terminal asset value and the \emph{quadratic variation} accumulated over $[0,T]$ . The payoff is then represented by the random variable $F(S_T, I_T)$, where $F(x,y)$ is an integrable function in the joint distribution of $S_T$ and $I_T$ for all the possible sets of spot variables. By the usual dynamic hedging argument we set up a portfolio that is long the contract, and short certain amounts of the underlying and of another variance dependent contract. By choosing the hedge ratios as to cancel the portfolio randomness, we argue that under no-arbitrage, for the given market price of risk, the portfolio process must earn the risk-free rate $r$. The time-$t$ value $V( S_t,  I_t, v_t, t)$ of the contract can be thus seen to satisfy the parabolic equation:

\begin{equation}\label{PDE}\frac{\partial V}{ \partial t}+ (r-d) S \frac{\partial V}{ \partial S}+  \alpha \frac{\partial V}{ \partial v} + v \frac{\partial V}{ \partial I}+   \frac{v S^2}{2} \frac{\partial^2 V}{ \partial S^2} +  \frac{\beta^2}{2} \frac{\partial^2 V}{\partial  v^2} + \rho \beta \sqrt{v} S \frac{\partial V}{\partial S \partial v}-rV=0.
\end{equation}

Pricing a cross asset-quadratic variation derivative $F(S_T, I_T)$ therefore amounts to solving the Cauchy free-boundary problem (\ref{PDE}) in $\mathbb R_+^3 \times [0,T]$ having terminal condition:

\begin{equation}\label{terminal}
V( S_T,  I_T, v_T, T)=F(S_T,I_T).
\end{equation}

\smallskip

This is the generalised version, in the price variable, of equation (13) of Sepp \cite{Sepp}, and the dual of the Fokker-Planck equation appearing in Lipton \cite{Lipton} and Fatone \emph{et al.} \cite{Fatone}.

As it happens when dealing with parabolic equations arising from financial modeling, results of existence/uniqueness of a solution may not be readily available from the standard theory of parabolic equations. Typically, this is because of two reasons: coefficients constraints are not met, or terminal conditions (payoffs) are not continuous. 
However, even if solvability remains an issue one has to live with\footnote{See for example Andersen and Piterbarg \cite{Andersen} on the non existence of moments in the Heston model.}, uniqueness of $V$ essentially carries over from the uniqueness of the underlying diffusion $X_t$, which is in turn enforced by assumptions (\ref{LL}) and (\ref{NB}).

Another interesting element of discussion is whether, and under which conditions, using the Feynman-Kac Theorem to link the discounted risk-neutral expectation of the payoff to the pricing equation can be considered to be equivalent to the no-arbitrage derivation\footnote{ As Lewis showed in \cite{Lewis} (ch. 9) this is not the case in models in which volatility explosions occurr with positive probability.}. This is a standard requirement in the literature; the following Proposition then motivates our assumptions on (D):

\begin{proposition}\label{uniqueness}Under assumptions (LL) and (NB) there exists at most one $C^{2,1}$ solution to problem (\ref{PDE})-(\ref{terminal}); if such a solution does exist, for  $x=(S_t,v_t,I_t)$ it is given by:
\begin{equation}\label{representation} V(x,t)=\E_{x,t}\left[e^{-r(T-t)} F(S_T,I_T) \right] .\end{equation}
\end{proposition}

Therefore under (\ref{LL}) and (\ref{NB}) pricing a claim of the form (\ref{payoff}) is a well posed problem, provided that (\ref{PDE}) is solvable.

We finally impose a few last growth constraints, this time directly on the solution $V$:
\begin{equation}\label{GC}\tag{GC}
\begin{array}{cc} V(S) < K_1(1+ S^{h_1}), & V(I) <K_2(1+I^{h_2})\end{array}
\end{equation}

\noindent for some $K_1(I), K_2(S) >0$, $h_1, h_2 \geq 0$.

The reason of this assumption is technical in nature and will allow to perform the necessary reductions while solving the equation. The classic theory of parabolic equations (Friedman \cite{Friedman}, \cite{Friedman2}) provides sufficient conditions on the problem itself under which (\ref{GC}) holds (\cite{Friedman2}, Th. 4.3, p. 147). However, it is difficult to give a comprehensive set of such assumptions in a financial setting, owing to the lack of the necessary regularity of many cases of interest: namely, superlinear growth of the coefficients of (D) or discontinuity of $F$. Alternatively, (\ref{GC}) can be checked case by case, for example by using estimates along the lines of those derived by Bergman {\em et al.} \cite{BGW} (Th. 1). 
In any case, this condition is easily seen to hold for most of the cases accounted in section 5 (see the appendix).


\section{Solution to the PDE}

We shall characterise the solution of the PDE by identifying a \emph{fundamental transform} for the problem, which loosely speaking is nothing but the characteristic function of the fundamental solution of the equation.  To do this we will apply the Fourier transform to (\ref{PDE}) with respect to both variables $I$ and $\log S$. Once a fundamental transform has been found, we can invert it on a suitable domain of $\mathbb C^2$ and then conclude from Proposition \ref{uniqueness} the existence of a unique price for $F$.

Let $V(t, S, v, I)$ be the solution of $(\ref{PDE})$ and consider the substitutions:

\begin{equation}\label{change2}
\left \{
\begin{array}{l}
 \tau=T-t
\\ x=\log S +(r-d)(T-t)
\\ W(x,y, v, \tau)= \left \{ \begin{array}{lc}e^{r(T-t)}V(S, I,v, t)  & \mbox{ if } y >0 \\
                                   0 & \mbox{ if } y \leq 0 . \end{array} \right.
\end{array}
\right.
\end{equation}

Equation (\ref{PDE}) can then be seen to be equivalent to the problem

\begin{equation}\label{PDE2}
\frac{v}{2} \left(\frac{\partial^2 W}{\partial x^2}-\frac{\partial W}{\partial x}+2 \frac{\partial W}{\partial y} \right)+\rho \beta \sqrt{v}  \frac{\partial W}{\partial x \partial v}+\alpha \frac{\partial W}{ \partial v}+ \frac{\beta^2}{2} \frac{\partial^2 W}{\partial v^2}= \frac{\partial W}{\partial \tau},
\end{equation}
with initial condition:
\begin{equation}\label{change}
W(x_0, y_0, v_0, 0)=\left \{
\begin{array}{lc}
 F(e^{x_0}, y_0) & \mbox{ if } y_0 >0 \\
 0  & \mbox{ if } y_0 \leq 0.
\end{array}
\right.
\end{equation}

For $(\eta, \omega) \in \mathbb C^2$, let the two-dimensional Fourier transform of $W(x,y, v, \tau)$ be:

\begin{equation}\hat W(\omega, \eta, v, \tau)= \int_{\mathbb R^2} e^{i x \omega +i y \eta} W(x,y,v,\tau)dx dy. \end{equation}

We denote derivatives by subscripts. Consider the transform $\widehat W_\tau$ of the time-to-maturity derivative of $W$.  By substituting (\ref{PDE2}) in the integral above and integrating by parts we find that
\begin{equation}\label{derivative}
\begin{array}{l c r}
\widehat W_\tau=\hat W_\tau , & \widehat W_x=-i \omega \hat W, &  \widehat W_{xx}= -\omega^2 \hat W \\
\widehat W_y=-i \eta \hat W,  & \widehat W_v= \hat W_v,  & \widehat W_{vv}= \hat W_{vv} \\
\widehat W_{xv}=-i \omega \hat W_v. & & \\
\end{array}
\end{equation}
provided that  $ e^{i \omega x} W(x)|_{-\infty} ^{+\infty}=e^{i \omega x} W_x(x)|_{-\infty} ^{+ \infty}= e^{i \eta y} W(y)|_{+ \infty}=0 $ holds true for some $\omega, \eta$. These relations are clear if we know $V$ to satisfy (\ref{GC}),  
 which then yields (\ref{derivative}) in a 2-strip $\Omega_1 = \{a_1< Im(\omega) <a_2, \, Im(\eta)>0 \} \subset  \mathbb C^2$.

Fourier-transforming both sides of (\ref{PDE2}) and substituting the above relations we have the fundamental PDE for $\hat W$:

\begin{equation}\label{PDE3}
 \frac{\beta ^2}{2}  \frac{\partial^2 \hat W}{\partial v^2}+  \frac{\partial \hat W}{\partial v }(\alpha- i \omega \sqrt{v} \rho \beta) -\frac{v}{2}(\omega^2-i \omega+ 2 i \eta) \hat W= \frac{\partial \hat W}{\partial \tau}.
 \end{equation}

A \emph{fundamental transform} $\hat H(\omega, \eta, v, \tau) $  for (\ref{PDE2}) is a solution to (\ref{PDE3}) such that $\hat H(\omega, \eta, v, 0)=1$.  Assume that such a solution exists:  $\hat H$ is nothing else than the (sign-shifted) characteristic function of the transition probability density associated with the process $(\log S_t, I_t)$, and it is thus a holomorphic function\footnote{See Lukacs \cite{Luckasz}, (Th. 7.1.1.). Since we are not confined to real arguments, we need not to consider analytical continuations around 0, and may instead develop around any point in whose neighbourhood $\hat H$ is holomorphic. This means that $\Omega_2$ will exists somewhere, even if in general it may not contain the real axis.} on a certain multi-strip $\Omega_2 \subset \mathbb C^2$.

Denote the Fourier transform of the payoff in the log-price and quadratic variation by $\hat F(\omega,\eta):=\hat W(\omega, \eta, v, 0)$, itself a holomorphic function on a third multi-strip $\Omega_3 \subset \mathbb C^2$. Since $ \hat F(\omega,\eta)$ \emph {does not depend on the variable v} we see that  the product $\hat H(\omega, \eta, v, \tau)  \hat F(\omega,\eta) $ is \emph{also a solution to (\ref{PDE3}) having
initial condition} $ \hat F(\omega,\eta)$. Therefore by taking the Fourier inverse of $\hat H(\omega, \eta, v, \tau)  \hat F(\omega,\eta) $  on a multi-line \begin{equation}\begin{array}{ll}\Sigma=\displaystyle{\{(\omega, \eta), \omega=s+ i k_1, \eta=t+i k_2, s,t \in \mathbb R \} \subset \Omega=\bigcap_{i=1}^3 \Omega_i}, & k_1, k_2 \in \mathbb R, \end{array}\end{equation}
and finally unwinding the variable change we are led to the solution of (\ref{PDE}):
\begin{align}\label{solution} V(S,I,v,t) =  \frac{e^{-r(T-t)}}{4 \pi^2}  \int_{ik_1- \infty}^{ik_1+\infty}\int_{ik_2- \infty}^{ik_2+\infty}  S^{-i \omega }e^{-i \omega (r-d)(T-t)}e^{-i \eta I}\hat H(\omega, \eta, v, T-t)  \hat F(\omega,\eta) d \omega d \eta.
\end{align}

\medskip

Finally, Proposition \ref{uniqueness} establishes that (\ref{solution}) is the unique price of $F(S_T,I_T)$. Of course, such an argument is meaningful provided that \emph{a common domain of holomorphy} $\Omega \subset \mathbb C^2$ \emph{of} $\hat H$ \emph{and} $\hat F$  \emph{actually exists}.

We summarise all of the above discussion in the following Proposition:

\begin{proposition}
Assume that the solution $X_t$ of (\ref{D}) is such that the dynamics for $v_t$ satisfy (\ref{LL}), (\ref{NB}), and that (\ref{GC}) holds. Further assume that $\Omega
\neq \emptyset$ and let $k_1, k_2 \in \R$ be such that:  \begin{equation}\Sigma=\{(\omega, \eta), \omega=s+ i k_1, \eta=t+i k_2, s,t \in \mathbb R \} \subset \Omega.\end{equation} If a fundamental transform $\hat H(\omega, \eta, v, \tau)$ can be found, the price of a claim $F$ written on $S_t$ and $I_t$ is given by equation (\ref{solution}).
\end{proposition}

This formula is completely general: in principle, under the given assumptions, it allows pricing under any stochastic volatility model.

Another attractive feature of equation (\ref{solution}) is that it allows us to separate,  by means of $\hat H$ and $\hat F$, the pricing information coming from the \emph{model} from that coming from the \emph{payoff}. This was one of the original main contributions of Lewis's work on  stochastic volatility models and is equally valid here. Changing the stochastic volatility or the function to be valued only requires changing the corresponding transform to be used in (\ref{solution}), and not the whole re-computation of the solution.

\subsection{Greeks}

The representation found also allows for a straightforward computation of the Greeks. Calling $J(\omega, \eta, v, \tau)$ the integrand in (\ref{solution}) and differentiating $V$ under integral sign we have that the Delta for the contract $F$ is:  
\begin{equation}\label{delta}
\Delta=\frac{\partial V}{\partial S}=-\frac{e^{-r(T-t)}}{4 \pi^2}  \int_{ik_1- \infty}^{ik_1+\infty}\int_{ik_2- \infty}^{ik_2+\infty} \frac{i \omega}{S} J(\omega, \eta, v, \tau) d \omega d \eta.
\end{equation}
 Likewise, the Gamma is seen to be given by:
\begin{equation}\label{gamma}
\Gamma=\frac{\partial^2 V}{\partial^2 S}=\frac{e^{-r(T-t)}}{4 \pi^2}  \int_{ik_1- \infty}^{ik_1+\infty}\int_{ik_2- \infty}^{ik_2+\infty} \frac{i \omega- \omega^2}{S^2} J(\omega, \eta, v, \tau) d \omega d \eta.
\end{equation}

The derivative $\Delta$ is one the two coefficients to be used in the hedge ratios yielding equation (\ref{PDE}). The sensitivity to the initial inatantaneous varinace $\partial V/ \partial v$ can be sometimes expressed in a similar fashion as (\ref{delta}) and (\ref{gamma}), for example in affine models. Clearly, the ability to fully hedge will depend also on the possibility to identify a fundamental set of securities for the market. However, the situation is no more general than that of a standard stochastic volatility set-up, because the diffusion (D) shows no more randomness than a model with an asset process and stochastic volatility only.

\section{Model-specific fundamental transforms}
 We analyse here in more detail the fundamental transforms of the Heston, the 3/2 and the GARCH models. The analytical tractability that characterises these in a standard stochastic volatility scenario carries over when realised volatility comes into the picture. Remarkably, the solution of (\ref{PDE3}) depends on the coefficient of the linear term as a parameter. As the variable $\eta$ appears only in such coefficient, the derivations are formally identical to that already present in literature, to which our formulae reduce when $\eta=0$.

Following are the transforms, together with their domain of holomorphy as functions of two complex variables. Being a Fourier integral, $\hat H$ is everywhere holomorphic in its domain of definition. Complex square roots are always understood to be the positive determination. As per our initial assumption, the parameters for the models already incorporate the market price of risk adjustment. For a sketch proof of the derivation of (\ref{HestonFundamental}) consult the appendix; a complete treatment is to be found in Lewis \cite{Lewis}.
\subsubsection*{Heston model}
Parameters are assumed to be constant. In such a case a lot is known about the model (Heston \cite{Heston}, CIR \cite{CIR}) and a fundamental transform can be easily obtained.
\begin{itemize}
\item Dynamics of the instantaneous variance:
 \begin{equation}d v_t= \kappa(\theta-v_t) dt+ \epsilon \sqrt{v_t} dW_t, \end{equation}
 with $\kappa, \theta, \epsilon >0$. Volatility explosions never occur; taking $2 \kappa \theta \geq \epsilon^2$ ensures that the 0 boundary is not attainable. Thus under these conditions assumption (NB) is met.
\item Fundamental transform:
\begin{equation}\label{HestonFundamental}
\begin{array}{l}
\displaystyle{\hat H(\omega, \eta, v, \tau)=\exp[C(\omega, \eta, \tau)+ vD(\omega, \eta, \tau)]} \\ \\
\displaystyle{ C(\omega, \eta, \tau)= \frac{\kappa \theta}{\epsilon^2}  \left( \tau( b(\omega)-d(\omega, \eta)) - 2 \log \left(   \frac{e^{-d(\omega, \eta)\tau}- c(\omega, \eta) }{1-c(\omega, \eta)}\right) \right)} \\ \\
\displaystyle{ D(\omega, \eta, \tau)=\frac{b(\omega)+d(\omega, \eta)}{\epsilon^2}\left( \frac{1-e^{d(\omega, \eta) \tau}}{1-c(\omega, \eta) e^{d(\omega, \eta) \tau}} \right) }   \\ \\
\displaystyle{c(\omega, \eta)=\frac{b(\omega)+d(\omega, \eta)}{b(\omega)-d(\omega, \eta)} } \\
\displaystyle{b(\omega)=\kappa+ i \epsilon \omega \rho } \\
\displaystyle{ d(\omega, \eta)=\sqrt{b(\omega)^2+\epsilon^2(\omega^2-i \omega +2 i \eta)}.  }
\end{array}
\end{equation}

The expression for $C$ uses the argument by Lord and Kahl \cite{Lord} to avoid discontinuity issues in the complex logarithm. The only singularities occur when $1- c(\omega, \eta) e^{d(\omega, \eta)\tau}=0$ causing divergence in both $C$ and $D$; hence the domain of holomorphy of $\hat H$ is $\mathbb C^2 \setminus \mathcal S_{\kappa, \epsilon, \rho, \tau}$ where
\begin{equation}\mathcal S_{\kappa, \epsilon, \rho, \tau}=\{(\omega, \eta) \in \mathbb C^2  | \;  e^{-d(\omega, \eta) \tau} = c(\omega, \eta) \}. \end{equation}

\end{itemize}

\subsubsection*{3/2 model}
We consider the general form as introduced by Lewis \cite{Lewis}:
\begin{itemize}
\item Dynamics of the instantaneous variance:
 \begin{equation}d v_t= \kappa(\theta v_t-v^2_t) dt+ \epsilon {v_t}^{3/2} dW_t. \end{equation}
   Whenever $2 \kappa \geq - \epsilon^2$ we have that $+ \infty$ is unattainable; the 0 boundary is natural for any choice of parameters. 
\item Fundamental transform:
\begin{equation}\label{32 Fundamental}
\begin{array}{l}
\displaystyle{\hat H(\omega, \eta, v, \tau)=\frac{\Gamma(\beta - \alpha)}{\Gamma(\beta)} X \left(  \frac{2 \kappa \theta}{\epsilon^2 v}, \kappa \theta \tau \right)^{\alpha} \phantom{}_1F_1 \left[ \alpha, \beta, -  X \left( \frac{2 \kappa \theta}{\epsilon^2 v}, \kappa \theta \tau \right) \right]}\\
\displaystyle{X(x,t)= \frac{x}{e^t-1}} \\
\displaystyle{\alpha(\omega, \eta)= c(\omega, \eta)-b(\omega)  } \\
\displaystyle{\beta(\omega, \eta)= 1+2 c(\omega, \eta) }\\
\displaystyle{b(\omega)=(\kappa+\epsilon^2/2+ i\omega \rho \epsilon)/\epsilon ^2 }\\
\displaystyle{c(\omega, \eta)=\sqrt{b(\omega)^2+  d(\omega, \eta)}  } \\
\displaystyle{ d(\omega, \eta)= 2(\omega^2-i \omega +2 i \eta)/ \epsilon^2}.
\end{array}
\end{equation}
$\phantom{}_1F_1(\alpha, \beta,z)$ is a confluent hypergeometric series and $\Gamma$ is the Euler's Gamma function. Since $\beta$ cannot be a negative integer the poles of $\Gamma$ are avoided, so that the domain of the transform is the whole $\mathbb C^2$.
\end{itemize}
\subsubsection*{GARCH model}
We only take in account a particular instance of this model, namely when the dynamics are simply those of a geometric Brownian motion with drift; we also assume  $\rho=0$. The case $\rho \neq 0$ can be obtained by a simple modification of the derivation in \cite{Lewis}.
\begin{itemize}
\item Dynamics of the instantaneous variance:
 \begin{equation}\label{GARCHFundamental}d v_t= \theta v_t dt+ \epsilon {v_t} dW_t, \end{equation}
 with $\epsilon >0$.  Clearly condition (\ref{NB}) is always met.
 \item Fundamental transform:
\begin{align}\hat H(&\omega, \eta, v, \tau) = \nonumber \\ & \frac{2^{\beta+1}} {d(\omega, \eta)^{\beta}}  \left[ \I_{\{\beta<0\}} \sum_{j=0} ^{[-\beta/2]} \frac{-\beta-2j}{j ! \Gamma(1-\beta-j)}K_{-\beta-2j}(d(\omega, \eta))e^{(\beta j+j^2)\epsilon^2 \tau/2} \right. \nonumber  \\ & + \left. \frac{1}{4 \pi^2} \int_0 ^{\infty} \Big| \Gamma \left(\frac{\beta+iz}{2} \right) \Big|^2 z \sinh(z \pi) K_{iz}(d(\omega, \eta))e^{-(\beta^2+z^2)\epsilon^2 \tau/8} dz \right] \nonumber \\ & \displaystyle{\beta=  2 \theta/\epsilon^2 -1 } \nonumber  \\
& \displaystyle{d(\omega, \eta)= 2  \sqrt{2 (\omega^2-i \omega +2 i \eta) v}/\epsilon}.  \nonumber \\
 \end{align}
Here $\I$ is the indicator function and $K_x$ the modified Bessel function of second kind.  By use of the appropriate series representation, $K_x$ can be extended to an entire function. So we see that $\hat H$ is a holomorphic on $\mathbb C^2 \setminus \{\omega^2-i \omega +2 i \eta \} $ whenever $2 \theta > \epsilon^2$, and it is everywhere analytical on $\mathbb C^2$ otherwise.
\end{itemize}

\section{Some joint asset/volatility derivatives}\label{derivatives}

We present here a list of European-style derivatives paying off a joint function of a terminal asset value and its realised variance or volatility. As explained in the introduction, these can all be considered as volatility-modified versions of classic payoffs, where the volatility factor reduces the initial price without effecting the payoff if the investor's volatility foresight happens to be correct. The target volatility option is one currently traded product of this kind. 

\subsubsection*{Target volatility option}

A target volatility call option is the option to buy a certain fractional amount of shares if the underlying is worth more than the strike price at maturity. Such amount is stochastic and  depends upon both the target parameter $\target$ set when writing the contract and the volatility realised by the asset in $[0,T]$. Under the independence hypothesis between the Brownian motion driving the underlying and the process for the instantaneous volatility, the value of an at-the-money Call TVO is approximately the Black-Scholes price of a call option of constant volatility $\target$ (see \cite{Torricelli}).
\begin{itemize}
\item  {\bf Payoff} \begin{equation}F(S_T, I_T)=\target \sqrt{ \frac{T}{\Vol}} (S_T - K)^+. \end{equation}
\item  {\bf Payoff transform in log-strike and quadratic variation}
   \begin{equation}\begin{array}{cr} \displaystyle{ \hat F(\omega,\eta) =  \target(1+i) \sqrt{ \frac{\pi T}{2 \eta}}  \frac{ K^{1+i \omega}}{(i \omega- \omega^2)}}&   \mbox{ for } Im(\omega) >1 \, , Im(\eta)>0  \end{array}.\end{equation}
\end{itemize}
As is the case for vanilla options, a target volatility put will have the same payoff transform as a call, but in the domain we will instead have Im$(\omega)<0$.

\subsubsection*{Double digital call}

 A derivative delivering at maturity a unit of cash if both the underlying asset and its realised variance at  $T$ are above two strike levels $K_1$ and $K_2$. In  practice we are adding a further strike threshold to a digital call option, so it is intuitively clear that this derivative must be priced less than it. 
\begin{itemize}
\item  {\bf Payoff}  \begin{equation}F(S_T,\Vol)= \I_{\left \{ S_T \geq K_1, \, \Vol / T \geq K_2 \right \}}. \end{equation}
\item  {\bf Payoff transform in log-strike and quadratic variation}
   \begin{equation}\begin{array}{cr} \displaystyle{ \hat F(\omega,\eta)= - \frac{K_1^{i \omega}e^{i T K_2 \eta}}{\omega \eta}}  &   \mbox{ for } Im(\omega) >0 \, , Im(\eta)>0.  \end{array} \end{equation}
\end{itemize}
Clearly any other Put/Call combination in the two variables can be imagined.

\subsubsection*{Volatilty capped call option}
As in the previous example we can cheapen the price of a European Call by adding the further constraint that the payoff is not triggered if the terminal realised volatility is not within an acceptable range.
\begin{itemize}
\item  {\bf Payoff}  \begin{equation}\label{capped}F(S_T,\Vol)= (S_T-K)^+ \I_{\{  K_1 \leq \sqrt{\Vol / T} \leq K_2 \}}. \end{equation}
\item  {\bf Payoff transform in log-strike and quadratic variation}
   \begin{equation}\begin{array}{cr} \displaystyle{\left( e^{i \eta K_1^2 T}- e^{i \eta K_2^2 T}\right) \frac{ K^{1+i \omega}}{( \omega +i \omega^2)\eta}}  &   \mbox{ for } Im(\omega) >1 \, , Im(\eta)>0.  \end{array} \end{equation}
\end{itemize}

A more natural version of this product is obtained by requiring that to get a positive payoff the volatility should never leave an interval $[K_1,K_2]$ at each given time $t<T$. The resulting derivative is a ``volatility version'' of a double barrier option; pricing it therefore amounts to solve a boundary-valued version of (\ref{PDE})-(\ref{terminal}). This escapes the pricing frame presented; however, such a payoff could be of interest for future research in a context of volatility path-dependent claims.

\subsubsection*{ Volatility struck call option}

This product gives the writer the option to buy an asset at maturity for a notional amount $N$, times the realised volatility of the underlying. The more the stock is subject to shocks, the less likely is the option to be triggered; hence an investor could enter this contract if he is expecting low volatility levels. Just like for a TVO, predictions about the future realised volatility $\sigma$ are reflected in setting the notional $N$. 


\begin{itemize}
\item  { \bf Payoff}  \begin{equation}\label{struck}F(S_T,\Vol)= \left (S_T - N \sqrt{\frac{\Vol}{T}} \right)^+. \end{equation}
\item  { \bf Payoff transform in log-strike and quadratic variation}

\begin{align}\hat F(\omega,\eta)=   & \left(\frac{ N }{\sqrt{T}} \right)^{1+i \eta} \Gamma\left( \frac{3+ i \omega}{2} \right)  \frac{(-i \eta)^{- 3/2-i \omega/2} } { i \omega- \omega^2 }  \nonumber \\ \nonumber \\ & \mbox{ for } 1<Im(\omega)<3, \, Im(\eta) >0   .\end{align}

\end{itemize}

\section{Numerical testing}

Computations for the payoffs introduced in section \ref{derivatives} have been carried out in a Heston model with parameters from Di Graziano and Torricelli \cite{Torricelli}, in order to compare the values obtained there with the new methodology. The underlying process for the variance is given by

\begin{equation}
\begin{array}{rr}
dv_t=\kappa(\theta-v_t)dt+  \eta \sqrt{v_t}dW_t,
\end{array}
\end{equation}
with:
\begin{equation}
\begin{array}{cccc}
\kappa=0.5, & \theta=0.2, & \eta=0.3, & v_{t_0}=0.2.
\end{array}
\end{equation}

For different state variables and sets of parameters defining the claims, we compare a MATLAB\textsuperscript{\textregistered} Monte Carlo simulation based on an Euler scheme with sampling from the log-normal distribution (Broadie and Kaya \cite{BroadieKaya}), against a MATHEMATICA\textsuperscript{\textregistered} implementation of (\ref{solution}). For the TVO, figures from the Laplace transform pricing method described in \cite{Torricelli} are provided. 
A comparison of the prices of the products introduced with their vanilla counterparts is also given; it is striking how much cheaper the new claims are. Nevertheless, under favourable volatility scenarios, they produce the same payoffs as their standard versions.


\begin{table}[h!]
\begin{center}
\caption{ \small{{\bf  $T=3$, $t=0$, $\bar \sigma=0.1$, $S_0=100,$ $r=d=\rho=0$. TVO valuation for different strikes.}}}
\vspace{10pt}
\begin{tabular}{| c | c | c | c | c |}
\hline
  $K$ & Laplace  & Monte Carlo  & PDE & Vanilla  \\ 
     & transform & simulation &  pricing & Call \\
  \hline
60    & 11.3919 &  11.3897  & 11.3909 &   40.0061
 \\
80    &  8.7301  &  8.7281   & 8.7299 &   20.7211 \\
 100 &     6.7416  &  6.7415   &  6.7415 &  6.9013 \\
 120    &   5.2672  &  5.2618 &  5.2672  &  1.4252 \\
\hline
\end{tabular}
\end{center}
{\footnotesize The Black-Scholes price of a Call option of constant volatility $\bar \sigma$ is given for comparison (see \cite{Torricelli}, sec. 7).}
\end{table}

\begin{table}[h!]
\begin{center}
\caption{ \small{{\bf  $T=5$, $t=2.5$, $r=0.08$, $d=0$, $\bar \sigma=0.1$, $S_t=100$, $K=85$, $I_t=0.46$. TVO prices for various correlations.}}}
\vspace{10pt}
\begin{tabular}{| c | c | c | c |}
\hline
 $\rho$  & Monte Carlo  & PDE pricing & Vanilla Call\\ 
  \hline
-0.8 &   10.3154 &10.3975 & 41.5145 \\
-0.4  & 9.9415  &  9.9505 & 41.3683\\
 0 &  9.4398 & 9.4549  & 41.1688  \\
 0.4 & 8.9645  &  8.9059 & 40.8992 \\
 0.8 & 8.3136   & 8.3025 & 40.5433 \\
\hline
\end{tabular}
\end{center}
{\footnotesize Here we compare to a European Call option in the Heston model having same parameters.}
\end{table}

\newpage

\begin{table}[h!]
\begin{center}
\caption{ \small{{\bf  $T=2.5$, $t=1$, $r=0.1$, $d=0.01$, $K_1=100$, $K_2=0.24$, $S_t=120$, $\rho=0.2$. Double digital call for different realised variance levels.}}}
\vspace{10pt}
\begin{tabular}{| c | c | c | c |}
\hline
  $I_t$ & Monte Carlo  & PDE pricing & Vanilla digital Call \\ 
  \hline
 0.2 & 0.0951   &  0.0943 & 0.5358\\
  0.3  &  0.2366  & 0.2426 & 0.5358 \\
  0.4 &   0.4393 &  0.4395 & 0.5358 \\
  0.5  &     0.5335  & 0.5330 & 0.5358 \\
 \hline
\end{tabular}
 \end{center}
\vspace{-.1cm}
{\footnotesize Note how the prices converge to that of a digital Call as $K_2$ becomes more likely to be hit.}
\end{table}

\begin{table}[h!]
\begin{center}
\caption{ \small{{\bf   $t=d=0$, $T=2$, $r=0.07$, $S_t=110$, $K=100$, $K_1=0.2$,  $\rho=-0.3$. Volatility capped call option prices for different values of $K_2$.}}}
\vspace{10pt}
\begin{tabular}{| c | c | c | c |}
\hline
 $K_2$  & Monte Carlo  & PDE pricing & Vanilla Call \\  
 \hline
0.35 &  7.7812   & 7.7743  &  37.2632 \\
0.4 &  16.3226 & 16.3006   &  37.2632  \\
0.45 &     25.1122 & 25.0732     &  37.2632  \\
 0.5 & 31.6069   & 31.5497     & 37.2632 \\
\hline
\end{tabular}
\end{center}
\vspace{-.1cm}
{\footnotesize The reference vanilla Call has same parameters as the volatility capped call. As the gap between $K_1$ and $K_2$ widens, the same behaviour as in Table 3 is shown.}
\end{table}

\begin{table}[h!]
\begin{center}
\caption{ \small{{\bf  $I_t=0.18$, $t=1$, $r=0.05$, $d=0.02$, $S_t=50$, $N=150$,  $\rho=-0.5$. Volatility struck call option prices for different maturities.}}}
\vspace{10pt}
\begin{tabular}{| c | c | c |} 
\hline
 $T$  & Monte Carlo  & PDE pricing \\ 
  \hline

 2 &  4.8291  & 4.8815  \\

  3 & 8.9873 & 8.9383  \\

 4 &  11.8885 & 11.9086   \\
 5 &  14.1661 & 14.2002    \\
\hline
\end{tabular}
\end{center}
\vspace{-.1cm}
\end{table}

\section{Conclusions}
In this paper we have explained possible reasons for the introduction in the markets of derivatives written jointly on an underlying asset and its accrued volatility. We have discussed the general problem of pricing European claims depending at maturity on such an asset and the total quadratic variation it exhibits. A pricing PDE has been derived, and a \emph{universal} model-dependent solution has been found and characterised in terms of the \emph{model} and the \emph{payoff}.  
Issues of uniqueness and existence of such a solution have been addressed and a detailed mathematical discussion of the domains of holomorphy of the involved functions has been carried out. Furthermore, we have provided an easy analytical  representation for the Greeks, and given formulae for specific models and payoffs.

Numerical tests support our main result. In addition, figures confirm the intuition that it is possible to conceive volatility modifications of liquid instruments, less expensive than the original product, but paying off the same amount in market situations an investor may want to exploit.

\section{Appendix}

\subsubsection*{Proof of Proposition \ref{uniqueness}}

The proof is an adaptation of Heath and Schweizer \cite{Heath} (Th. 1), the existence of a solution being assumed. See also Friedman \cite{Friedman2} (ch. 5-6).

Let $D_n$ be a family of smooth bounded domains invading $\mathbb R^3_+$, and assume $V(x,t)$ is a $C^{2,1}(\mathbb R_+^3 \times [0,T))$ solution to (\ref{PDE})-(\ref{terminal}). Let $\tau_n= \{\inf t | X_t \notin \overline D_n \}$. By It\^{o}'s formula  it can be readily seen that if $X_t$ is a solution to  (\ref{D}) then for all $x \in D_n, t \leq T$,
\begin{equation}\label{boundary}V_n(x,t)=\E_{x,t}\left[e^{-r(T-t)}V(X_{T \wedge \tau_n}, T \wedge \tau_n ) \right]
\end{equation}
is a solution to the differential problem $(\ref{PDE})$ with boundary condition $V_n(x,t)=V(x,t)$, $x \in \partial D_n$, and terminal condition $V(X_{T \wedge \tau_n}, T \wedge \tau_n )$; furthermore, it is the only solution there by the weak maximum principle. By (NB) and (LL) we have that the probabilistic family yielding (\ref{boundary}) is strongly Markovian, so conditioning (\ref{representation}) at time $\tau_n$ and then taking the expectation shows that $V(x,t)$ coincides with $V_n$ on $D_n$ for all $n$. Hence, $V$ satisfies $(\ref{PDE})-(\ref{terminal})$ on the whole space. Finally, again (LL) and (NB) imply (see for example Feller \cite{Feller1}, Kunita \cite{Kunita}) that $v_t$, hence $X_t$, is (weakly) unique and finite almost surely, and this proves the claim.

\subsubsection*{Condition (GC) for the payoffs of section 5}

Let $V(S,I)=\E_{x,t} \left[e^{-r(T-t)}F(S,I) \right]$ where $F$ is a payoff from section 5. For a double digital call we have $F(S,I) \leq 1$ so that $V(S,I) \leq 1$ and (\ref{GC}) is trivially verified. If $F$ is a volatility capped call then $\forall I$:
\begin{equation} V(S,I) \leq \E_{x,t} \left[e^{-r(T-t)} (S_T-K)^+ \right] \leq S,
\end{equation}
where the last inequality follows at once from \cite{BGW} (Th.1) and its generalisation on page 1600. Also, by arguing that the {\em no-crossing} property (Lemma p. 1577) for $S_t$ must also hold for the augmented diffusion (D),
 we see that Theorem 1 
 can be extended to a payoff $F(S,I)$. In particular if $F$ is given by (\ref{struck}) we see that:
\begin{equation}\frac{\partial V(S,I)} { \partial S} \leq \sup_S \frac{\partial F(S,I)} { \partial S}=1.    \end{equation}

Hence $V(S,I)  \leq S$, $\forall S,I$.
 Similarly for a TVO we have:
 \begin{equation}\frac{\partial V(S,I)} { \partial S} \leq \target \sqrt{T/I}, \end{equation}
which implies
$V(S,I) \leq C_1 S$
for some $C_1 >0$, because $V$ is bounded around $I=0$.

\subsubsection*{Proof of equation (\ref{HestonFundamental})}

 We make the ansatz: \begin{equation}\hat H(\omega, \eta, v, \tau)=\exp[C(\omega, \eta, \tau)+ vD(\omega, \eta, \tau)]\end{equation}
and substitute this in (\ref{PDE3}) with parameters from the Heston model. One obtains the decoupled ODEs for  $C(\tau)$ and $D(\tau)$:
\begin{equation}
\begin{array}{l}
C'=\kappa \theta D \\
\displaystyle{D'=\frac{\epsilon ^2}{2}D^2 +D(\kappa  +i \epsilon \rho \omega)-\frac{1}{2}(\omega^2-i \omega+2 i \eta)}.
\end{array}
\end{equation}
The Riccati equation for $D$ is solved by switching to an associated linear second order ODE for its logarithmic derivative. $C$ is then found by direct integration.




\subsubsection*{Coniditon (NB) for the accounted models}

For the GARCH model the result is trivial. 
   The most convincing way of checking (NB) for the other models described is using Feller's explosion test (Feller, \cite{Feller2}).  In our case, the \emph{scale function} for the variance dynamics in the Heston model is:
\begin{equation}p(x)=\frac{2}{\epsilon^2}\int_1^x u^{-2 \kappa \theta/ \epsilon^2} e^{\frac{ 2 \kappa}{ \epsilon^2}(u-1)} \left(\int_1^u z^{2 \kappa \theta/ \epsilon^2-1} e^{- \frac{2 \kappa }{\epsilon^2}(z-1)} dz \right) du\end{equation}
A necessary and sufficient condition for the process to attain $0$ or $+\infty$ is that $p(0),$ $p(+\infty)< + \infty$. As $ u \rightarrow + \infty$ the integrand is exponentially divergent, whereas convergence in $0$ is happens if and only if $2 \kappa \theta < \epsilon^2$. For the 3/2 model we have instead:
\begin{equation}p(x)=\frac{2}{\epsilon^2}\int_1^x u^{2 \kappa / \epsilon^2} e^{ \frac{4 \kappa \theta }{ \epsilon^2}(1/u-1)} \left(\int_1^u z^{-2 \kappa/ \epsilon^2-3} e^{-\frac{4 \kappa \theta }{  \epsilon^2}(1/z-1)} dz\right) du . \end{equation}
so that this time it is $p(0)= + \infty$ for any choice of parameters, and $p(+\infty)<+ \infty$ if and only if $2 \kappa  < - \epsilon^2$.

\bibliographystyle{plain}

\end{document}